%% file: output.tex
\documentclass[a4paper,UKenglish,cleveref, autoref, thm-restate]{lipics-v2021}

\pdfoutput=1 
\hideLIPIcs  

\usepackage[unicode]{hyperref}
\usepackage[utf8]{inputenc}

\usepackage{amssymb}        
\usepackage{latexsym}
\usepackage{array}
\usepackage{anyfontsize}

\usepackage{algorithm}
\usepackage[noend]{algpseudocode}
\usepackage{amsmath} 
\usepackage{verbatim} 

\title{Rectangle Tiling Binary Arrays} 

\author{Pratik Ghosal}{Indian Institute of Technology, Palakkad, India}{pratik@iitpkd.ac.in}{https://orcid.org/0000-0002-4416-5160}{}

\author{Syed Mohammad Meesum}{Krea University, India}{meesum.syed@krea.edu.in}{https://orcid.org/0000-0002-1771-403X}{}

\author{Katarzyna Paluch}{University of Wroc\l aw, Wroc\l aw, Poland}{abraka@cs.uni.wroc.pl}{https://orcid.org/0000-0002-7504-6340}{}

\authorrunning{P Ghosal et al.} 

\Copyright{Pratik Ghosal, Syed Mohammad Meesum and Katarzyna Paluch} 



\category{APPROX} 

\relatedversion{https://arxiv.org/pdf/2007.14142} 

\funding{This work was partially funded by Polish National Science Center grant UMO-2018/29/B/ST6/02633. }

\nolinenumbers 

\EventEditors{Amit Kumar and Noga Ron-Zewi}
\EventNoEds{2}
\EventLongTitle{Approximation, Randomization, and Combinatorial Optimization. Algorithms and Techniques (APPROX/RANDOM 2024)}
\EventShortTitle{\mbox{\scriptsize APPROX/RANDOM 2024}}
\EventAcronym{APPROX/RANDOM}
\EventYear{2024}
\EventDate{August 28--30, 2024}
\EventLocation{London School of Economics, London, UK}
\EventLogo{}
\SeriesVolume{317}
\ArticleNo{28}

\usepackage{xspace} 
\newcommand{\RTILE}{{\sc RTile}\xspace}
\newcommand{\DRTILE}{{\sc DRTile}\xspace}
\newcommand{\apxfac}{\frac{3}{2}}
\newcommand{\RPACK}{{\sc RPack}\xspace}
\usepackage{mathtools}

\DeclarePairedDelimiter{\ceil}{\lceil}{\rceil}

\newcommand{\dowod}{\noindent{\bf Proof.~}}
\newcommand{\koniec}{\hfill $\Box$\\[.1ex]}
\newtheorem{fact}{Fact}
\newtheorem{convention}{Convention}

\usepackage{tikz}
\usetikzlibrary{decorations.markings}
\usetikzlibrary{shapes.geometric}
\pgfdeclarelayer{edgelayer}
\pgfdeclarelayer{nodelayer}
\pgfsetlayers{edgelayer,nodelayer,main}

\tikzstyle{none}=[inner sep=0pt]

\tikzstyle{rect}=[rectangle,fill=black,draw=black, scale=0.7]
\tikzstyle{dott}=[circle,fill=black,draw=black,scale=0.25]

\tikzstyle{simple}=[-,draw=Black,line width=2.000]
\tikzstyle{arrow}=[-,draw=Black,postaction={decorate},decoration={markings,mark=at position .5 with {\arrow{>}}},line width=2.000]
\tikzstyle{tick}=[-,draw=Black,postaction={decorate},decoration={markings,mark=at position .5 with {\draw (0,-0.1) -- (0,0.1);}},line width=2.000]

\begin{document}

\maketitle


\sloppy

\input{0.abstract}
\input{1.introduction}
\input{2.Preliminaries.tex}
\input{3.two_dimensional.tex}

\input{4.Tightness.tex}

\input{5.drtile.tex}
\input{6.multi.tex}

\bibliography{output}


\end{document}

%% file: 0.abstract.tex
\begin{abstract}
The problem of rectangle tiling binary arrays is defined as follows. Given an $n \times n$ array $A$ of zeros and ones and a natural number $p$, our task is to partition $A$ into at most $p$ rectangular tiles, so that the maximal weight of a tile is minimized. A tile is any rectangular subarray of $A$. The weight of a tile is the sum of elements that fall within it. We present a linear $(O(n^2))$ time $(\frac{3}{2}+\frac{p^2}{w(A)})$-approximation
algorithm (where $\frac{p^2}{w(A)} < \frac{1}{2}$) for this problem, where $w(A)$ denotes the weight of the whole array $A$. This improves on the previously known approximation with the ratio $2$.

The result is best possible in the following sense.
The algorithm employs the lower bound of $L=\lceil \frac{w(A)}{p} \rceil$,  
 which is the only known and used  bound on the optimum   in all algorithms for rectangle tiling.
We prove that a better approximation factor for the binary \RTILE cannot be achieved using  $L$, because there exist arrays, whose every partition
contains a tile with weight  at least $(\frac{3}{2}+\frac{p^2}{w(A)})L$. 
We also consider the dual problem of rectangle tiling  for binary arrays, where we are given an upper bound  on the weight of the tiles, and we have to cover the array $A$ with the minimum number of non-overlapping tiles. Both problems have natural extensions to $d$-dimensional
 versions, for which we provide analogous results.
\end{abstract}

%% file: 1.introduction.tex
\section{Introduction}
In this paper  we study several variants of the rectangle tiling problem. These problems belong to a very wide class of discrete optimization tiling problems. As an input, we are given a two-dimensional array $A[1...n,1...n]$, where each cell $A[i,j]$ has a non-negative weight. 

\textbf{\RTILE:} Given a two-dimensional array $A$ of size $n \times n$ and a natural number $p$, we partition $A$ into at most $p$ rectangular subarrays, called {\em tiles}, so that the maximum weight of any tile is minimized. In other words, we have to cover  $A$ with tiles such that no two tiles overlap, while minimizing the weight of any tile. The weight of a tile is the sum of  the elements that fall within it. 

\textbf{\DRTILE :} A natural variant of  \RTILE  is called the \DRTILE problem. The \DRTILE problem is a dual of the  \RTILE problem, where we are given an upper bound $W$ on the weight of the tiles, and we have to cover the array $A$ with the minimum number of non-overlapping tiles.    

 These two problems have a natural extension to  $d$ dimensions. Here the input is a $d$-dimensional array $A$ of size $n$ in each dimension and we have to partition $A$ into non-overlapping $d$-dimensional tiles so that the optimality criterion of the \RTILE/\DRTILE problem is satisfied.   
 
 In this paper we  consider a special case of the \RTILE/\DRTILE problem, where each cell has a binary weight, i.e., the weight of any cell is either $0$ or $1$. We  extend our approach to solve the $d$-dimensional binary \RTILE/\DRTILE problem.

\textbf{Motivation:} The \RTILE / \DRTILE problem is a  general problem in combinatorial optimization that has a wide variety  of applications in real life. These include load balancing in parallel computing environments, video compression, data partitioning, database mining, and building equisum histogram on two or more attributes. A detailed description of the practical applications of \RTILE / \DRTILE problem can be found in \cite{BermanDMR01, KhannaMP98, MuthukrishnanPS99}.

\textbf{Related Work:}  Both the \RTILE and \DRTILE problems can be solved in polynomial time when the array is one-dimensional. 
The \RTILE problem can be solved using dynamic programming in time $O(np)$. For any fixed $\epsilon < 1$, the best known algorithm has the running time $O(\min\{n + p^{1+\epsilon}, n \log n\})$ \cite{KhannaMS97}.
An extensive survey on the \RTILE problem in one-dimension can be found in \cite{KhannaMS97}. On the other hand, the \DRTILE problem in one dimension can be solved using a greedy algorithm in linear time.

Both the \RTILE and \DRTILE problems have been proven to be NP-hard \cite{KhannaMP98}. Grigni and Manne \cite{GrigniM96} proved that optimal $p \times p$ tiling (which is a restricted variant of the \RTILE problem) is NP-hard even when the cell weight is binary. Charikar et al. \cite{charikar} showed this problem to be APX-hard and  NP-hard to approximate within a factor of $2$. Khanna et al. \cite{KhannaMP98} proved  the \RTILE problem to be NP-hard to achieve a $\frac{5}{4}$-approximation. Recently Głuch and Loryś \cite{GluchL17} have improved the lower bound of the \RTILE problem to $\frac{4}{3}$. It is not known whether the binary \RTILE 
is solvable in polynomial time or NP-hard.  Khanna et al. \cite{KhannaMP98} gave the first approximation algorithm for the \RTILE problem with the ratio $\frac{5}{2}$. The approximation ratio was improved to $\frac{7}{3}$ independently by Sharp \cite{sharp1999tiling} and Loryś and Paluch \cite{LorysP00}.  Loryś and Paluch \cite{LorysP03} gave a $\frac{9}{4}$-approximation algorithm for this problem. Berman et al. \cite{BermanDMR01} improved the approximation ratio to $\frac{11}{5}$. Finally, Paluch \cite{Paluch04} gave a $\frac{17}{8}$-approximation for this problem and also proved that the approximation ratio is tight with respect to  the used lower bound. As far as the \DRTILE problem is concerned, Khanna et al. \cite{KhannaMP98} gave an $O(n^5)$-time $4$-approximation algorithm using the Hierarchical Binary Tiling (HBT) technique. They improved the approximation ratio to $2$ using a modified version of the HBT  technique, but the running time of this algorithm is very high making the algorithm less practical. Loryś and Paluch \cite{LorysP00} also gave a $4$-approximation for the \DRTILE problem while improving the running time to linear.

The $d$-dimensional version of this problem was introduced by Smith and Suri \cite{SmithS00}. They gave a $\frac{d+3}{2}$-approximation algorithm that runs in time $O(n^d + p\log n^d)$. Sharp \cite{sharp1999tiling} improved the approximation ratio to $\frac{d^2 + 2d -1}{2d - 1}$ that runs in time   $O(n^d + 2^dp \log n^d)$. Paluch \cite{Paluch06} gave a $\frac{d+2}{2}$-approximation algorithm while matching the previous running time. She also proved that the ratio is tight with respect to the known lower bound of the problem.

\RPACK is an extensively studied variant of rectangle tiling, in which we are given a set of axis-parallel weighted rectangles in a $n \times n$ grid, and the goal is to find at most $k$ disjoint rectangles of largest weight. Khanna et al. \cite{KhannaMP98} proved that this problem is NP-hard even when each rectangle  intersects  at most three other rectangles. They gave an $O(\log n)$-approximation algorithm for \RPACK that runs in $O(n^2p\log n)$ time. In \cite{BermanDMR01} Berman et al.  considered the multi-dimensional version of this problem. The dual of \RPACK is known to be NP-hard even when we are interested in finding a sub-set of disjoint rectangles with a total weight equal to at least some given $w$.  Du et al. \cite{DuEGL09} considered a min-max version of \RTILE, where the weight of each tile cannot be smaller than the given lower bound and  the aim is to minimize the maximum weight of a tile. They \cite{DuEGL09} gave a $5$-approximation algorithm for this problem and Berman and Raskhodnikova \cite{BermanR14} improved the approximation factor  to $4$ and the approximation ratio of the binary variant to $3$.

\textbf{Previous Work:} The binary version of the \RTILE problem has also been studied. Khanna et al. \cite{KhannaMP98} gave a $\frac{9}{4}$-approximation for the binary \RTILE problem. Loryś and Paluch \cite{LorysP00} and Berman et al. \cite{BermanDMR01} independently improved the approximation ratio for binary \RTILE  to $2$.

\textbf{Our Results:} We improve the approximation ratio of the binary \RTILE problem to $\frac{3}{2}+\frac{p^2}{w(A)}$,  where $w(A)$ denotes the number of ones in $A$.  For the arrays $A$ satisfying  $\frac{p^2}{w(A)} \approx 0$, it implies that   the approximation ratio of the algorithm amounts to $\frac{3}{2}$. 
The running time of our algorithm is linear ($O(n^2)$). The approximation is best possible in the following sense. The algorithm employs the lower bound of $L=\lceil \frac{w(A)}{p} \rceil$, which is the only known and used  bound on the optimum  in all algorithms for rectangle tiling.
We prove that a better approximation factor for the binary \RTILE cannot be achieved using $L$, because there exist arrays, whose every partition
contains a tile of weight at least $(\frac{3}{2}+\frac{p^2}{w(A)})L$.

The general approach to solving this problem is to some extent similar  to the approach of \cite{Paluch04}. The found tiling is also hierarchical and we use the notions of {\em boundaries} and {\em types of columns/subarrays} as well as we apply linear programming in a non-standard way. However, in the present paper the types of subarrays are organized in a somewhat different manner. In particular,
the idea of {\em shadows} is new. To compute the desired partition of $A$ into tiles, we only check a small number of tilings of simply defined subarrays. The subarrays are identified with the help of so called {\em boundaries} and their {\em shadows}, which, roughly speaking, designate parts of $A$ tileable in a certain manner and having a weight greater than $\frac{3}{2}L$. To prove the tileability of subarrays composed of multiple simpler subarrays we employ linear programming. 
Its application here differs from the one in \cite{Paluch04} in that each dimension is treated completely symmetrically and thus more "globally" and in the method of showing the feasibility of dual programs. 
We show that the binary \DRTILE problem can be approximated by reducing it to the binary \RTILE problem.  As for the $d$-dimensional binary \RTILE problem, the algorithm for the $2$-dimensional binary \RTILE problem can be extended to obtain an approximation for the $d$-dimensional binary \RTILE problem. The same approximation ratio for the d-dimensional binary \DRTILE problem can also be found analogously.

\textbf{Organization:} In Section \ref{section2}, we recall the necessary definitions. In Section $\ref{boundariessection}$, we revisit the definition of a boundary and introduce shadows of a boundary. In Section \ref{easyalgorithm}, we assume that $w(A) \gg p^2$ and present a $\frac{3}{2}$-approximation algorithm for the \RTILE problem.
The goal of this section is also to introduce the methods needed for the approximation of the binary \RTILE more gradually, without obscuring the presentation with many technical aspects.
In Section \ref{hardalgoriithm} we present a $(\frac{3}{2} + \frac{p^2}{w(A)})$-approximation algorithm for the general case (which, in particular, applies also when $\frac{p^2}{w(A)}$ is not negligible). This approximation is achieved by applying only small modifications to the approach described in Section \ref{easyalgorithm}. In Section \ref{tightness}, we show that the approximation factor we obtain for the \RTILE problem is tight with respect to the known lower bound. Due to space constraints, we moved the following sections to the appendix. Section \ref{drtile} contains our result on the \DRTILE problem. We conclude by presenting an approximation algorithm for the multi-dimentional \RTILE problem in Section \ref{multi}.

%% file: 2.Preliminaries.tex
\section{Preliminaries} \label{section2}
Let $A$ be a two-dimensional array of size $n \times n$, where each of its elements belongs to the set $\{0,1\}$. Given $A$ and a natural number $p$, we want to partition $A$ into $p$ rectangular subarrays, called {\em tiles} so that the maximal weight of a tile is minimized. The weight of a tile $T$, denoted  $w(T)$, is  the sum of  elements within $T$. $w(A)$ denotes the weight of the whole array $A$. Since any array element is either equal to $0$ or $1$, $w(A)$ amounts to the number of $1$s  in $A$.

First, notice that the problem is trivial when $p \geq w(A)$. Assume then that $p < w(A)$.
Clearly, the maximal weight of a tile cannot be smaller than $\frac{w(A)}{p}$.
Consequently, $L=\lceil \frac{w(A)}{p}\rceil$  is a lower bound on the value of the optimal solution to the \RTILE problem. 

Thus to design an $\alpha$-approximation algorithm for the \RTILE problem, it suffices to demonstrate the method of partitioning $A$ into $p$ tiles
such that the weight of each tile does not surpass $\alpha L$.

The number $p$ of allowed tiles is linked to the weight of the array $A$ in the following manner.

\begin{fact} \label{nrtiles}
Let $w(A)$ and $L$ be as defined above. Then, $p \geq  \lceil \frac{w(A)}{L} \rceil$. 
\end{fact}

The proof directly  follows from the assumption that $L = \ceil{\frac{w(A)}{p}}$


\begin{definition}
An array $A$ is said to be $f$-partitioned if it is partitioned into rectangular tiles such that the weight of any tile does not exceed $f$. 
\end{definition}

We denote by $A[i]$ the $i$-th column of $A$,
by $A[i..j]$  a subarray of $A$ consisting of columns $i, i+1, \ldots, j$. Thus $A^T[i]$ denotes the $i$-th row of $A$ and
$A^T[i..j]$  a subarray of $A$ consisting of rows $i, i+1, \ldots, j$.

%% file: 3.Two_dimensional.tex
\section{The Boundaries and Their Shadows} \label{boundariessection}
Let us assume that we want to design an $\alpha$-approximation algorithm for the \RTILE problem. Hence the weight of any tile must not exceed $\alpha L$. In other words, we want to obtain an $\alpha L$-partitioning for $A$. 

To help find such a partitioning we are going to make use of a sequence of (vertical) \emph{boundaries} and their \emph{shadows}. The vertical boundaries and shadows of array $A$ are defined iteratively below.
Each boundary and each shadow is a distinct column of $A$. The $i$-th boundary of $A$ is denoted as $B_i=A[b_i]$, i.e., $B_i$ is the $b_i$-th column of $A$ (or equivalently, $B_i=A[k]$, where $k=b_i$). Similarly, the $i$-th shadow of $A$ is denoted as  $B'_i=A[b'_i]$. The number of boundaries and their shadows depends on the weight and structure of $A$. 
 The shadow $B'_i=A[b'_i]$  is equal to either $B_i$ or the column succeeding $B_i$, i.e. either $b'_i=b_i$, or $b'_i=b_i+1$.  For each boundary $B_i$ we define its {\em type} -  we say that boundary $B_i$ is { of type} $j$,  denoted as $t(B_i)=j$, if its weight satisfies  $\lfloor \frac{w(B_i)}{\alpha L} \rfloor =j-1$. 

The ideas behind boundaries and shadows  are as follows. The first vertical boundary $B_1$ indicates simply which part of the array consisting of successive columns starting from the leftmost,
 exceeds $\alpha L$. This means that such a subarray cannot be covered with one tile. However, the subarray $A[1..b_1-1]$  ending on column $b_1-1$ can form one tile, because its weight is not greater than $\alpha L$. For $i>1$ the $i$-th boundary $B_i=A[b_i]$ is established in the following way. We distinguish two cases: $(i)\  B_{i-1}= B'_{i-1}$ and $(ii)\  B_{i-1}\neq B'_{i-1}$. Let us first consider case $(i)$. Suppose that $t(B_{i-1})=j$. Any boundary of type $j$ can be $\alpha L$-partitioned (horizontally) into $j$ tiles. We  check how far to the right we are able to extend one of such partitions. Thus, to identify the $i$th boundary $B_i$,  we find $b_i$ such that the subarray $A[b'_{i-1} .. b_i-1]$ can be $\alpha L$-partitioned horizontally into $j$ tiles and the subarray  $A[b'_{i-1} .. b_i]$ cannot. When $(ii) \  B_{i-1}\neq B'_{i-1}$, to identify the $i$th boundary $B_i$,  we proceed in the same way as with the first boundary $B_1$, i.e., we find $b_i$ such that the subarray $A[b'_{i-1} .. b_i-1]$ can be $\alpha L$-partitioned horizontally into $1$ tile and the subarray  $A[b'_{i-1} .. b_i]$ cannot.

As for the $i$-th shadow $B'_i$ we put it in the same column as the boundary $B_i$ if the subarray $A[b'_{i-1}..b_i]$ cannot be $\alpha L$-partitioned into $t(B_i)$ tiles and otherwise, we put it just behind $B_i$ - in column $b_i+1$. Notice that in the case of a shadow we check the tileability into $t(B_i)$ tiles and not $t(B_{i-1})$. Also, we observe that $B_i \neq B'_i$ can happen only when $t(B_i)>t(B_{i-1})$ or $B_{i-1} \neq B'_{i-1}$. If $B_i \neq B'_i$, then it means, as we later prove, that the subarray $A[1..b_i]$ is rather easy to partition and we could in fact tile it with a proper number of tiles and start the process of tiling anew with the subarray $A[b'_i .. n]$.

We now give a formal definition of a sequence of (vertical) \emph{boundaries} of $A$ and their \emph{shadows}.  
For technical reasons we introduce a column $A[0]$ to array $A$.

    
    
   

            

\begin{definition}~ \label{def:boundary}
A boundary \( B_i \) is of type \( j \), denoted as \( t(B_i) = j \), if its weight satisfies \( \left\lfloor \frac{w(B_i)}{\alpha L} \right\rfloor = j - 1 \). Based on that, the boundaries and their shadows are defined as follows:
\begin{enumerate}
    \item $B[0]=A[0], B'_0=A[1]$, thus $b_0=0$ and $b'_0=1$,
    
    \item \textbf{$i$-th boundary $B_{i}:$} 
    \begin{enumerate}
        \item \label{3a} If $B_{i-1} = B'_{i-1}$, then $B_{i} = A[b_{i}]$ iff \\
         $A[b_{i-1}..b_{i}-1]$ can be $\alpha L$-partitioned horizontally into $t(B_{i-1})$ tiles and  $A[b_{i-1}..b_{i}]$ cannot. 
    
        \item \label{3b} If $B_{i-1} \neq B'_{i-1}$, then $B_{i} = A[b_{i}]$ iff \\
    $w(A[b_{i-1}+1..b_{i}-1])\leq \alpha L$  and $w(A[b_{i-1}+1..b_{i}])>\alpha L$.
   
\end{enumerate}

\item \textbf{$i$-th Shadow $B'_i$:} Let $t(B_{i}) = j$. \\
 \label{4b} $B'_{i} = B_{i}$ iff $A[b'_{i-1}..b_{i}]$ cannot be $\alpha L$-partitioned horizontally into $j$ tiles.
            
\end{enumerate}
\end{definition}

The horizontal boundaries are defined analogously. To illustrate the notion of boundaries and shadows let us consider a few examples. \\

\begin{figure}
\centering
\begin{tikzpicture}
[scale=.25,colorstyle/.style={circle, draw=black!100,fill=black!100, thick, inner sep=0pt, minimum size=4mm}]
    \draw[thick] (0,0)--(24,0);
    \draw[thick] (0,14)--(24,14);
    \draw[thick] (0,0)--(0,14);
    \draw[thick] (24,0)--(24,14);
    
    \draw[thick] (3,0)--(3,14);

    \draw[thick] (6,0)--(6,14);

    \draw[thick] (9,0)--(9,14);

    \draw[thick] (12,0)--(12,14);

    \draw[thick] (15,0)--(15,14);
    \node at (18,7) []{$\ldots\ldots$};
    \draw[thick] (21,0)--(21,14);
    
    \draw[thick, <-, > = latex] (0,16.5)--(4,16.5);
    \draw[thick, ->, > = latex] (8,16.5)--(12,16.5);
    \draw[thick, <-, > = latex] (0,15)--(2.5,15);
    \draw[thick, ->, > = latex] (6.5,15)--(9,15);
    \draw[fill = gray, opacity =0.3] (9,0) rectangle (12,14);

    \node at (4.5,15)[]{$\leq \frac{3}{2}L$};
    \node at (6,16.5)[]{$> \frac{3}{2}L$};

    \node at (1.5,-1){$A[1]$};
    \node at (4.5,-1){$A[2]$};
    \node at (7.5,-1){$A[3]$};
    \node at (10.5,-1){$A[4]$};
    \node at (13.5,-1){$A[5]$};
    \node at (22.5,-1){$A[n]$};
\end{tikzpicture}
\caption{An array with  column $A[4]$ as the only boundary}
\label{easy_boundary}
\end{figure}
\begin{example}\label{example1}
Array $A$  has only one vertical boundary $B_1=A[4]$ of type $1$. 

\noindent This means that the total weight of the first $3$ columns does not exceed $\alpha L$, i.e., $w(A[1..3]) \leq \alpha L$, and
the weight of the subarray consisting of columns $1 \ldots 4$ does - $w(A[1..4]) > \alpha L$. Since $t(B_1)=1$, by the definition, the weight
of $B_1=A[4]$ is not greater than $\alpha L$ and the shadow $B'_1$ of $B_1$ coincides with $B_1$. 
Since $A$ has only one boundary, it means that the weight of the subarray consisting of all columns except for the first $3$ is not greater than $\alpha L$, i.e., $w(A[4..n]) \leq \alpha L$. \\
\end{example}

\begin{example}\label{example2}
Array $A$  has only one vertical boundary $B_1=A[4]$ of type $2$ and $B'_1=A[4]$. 

\noindent Exactly as in the example above, we have $w(A[1..3]) \leq \alpha L$ and  $w(A[1..4]) > \alpha L$. 
The weight of $B_1$ satisfies: $3L \geq w(A[4]) > \alpha L$, because $t(B_1)=2$.
By the fact that $B'_1=B_1$, we know that the horizontal partition of $A[1..4]$ into $2$ tiles of weight not surpassing $\alpha L$
is impossible.  Since $A$ has only one boundary, we obtain that $A[4..n]$ can be partitioned into $t(B_1)=2$ tiles. \\

\end{example}

\begin{example}\label{example3}
Array $A$  has only one vertical boundary $B_1=A[4]$ of type $2$ and $B'_1=A[5]$. 

\noindent Again, we have $w(A[1..3]) \leq \alpha L$ and  $w(A[1..4]) > \alpha L$. This time, however, $B'_1 \neq B_1$, therefore  $A[1..4]$ can be 
partitioned into $2$ horizontal tiles with weight $\alpha L$ at most. Since $A$ has only one boundary and $B'_1 \neq B_1$, we have that
 $w(A[5..n]) \leq \alpha L$.
\end{example}

\begin{restatable}{lemma}{lemmaone} \label{tiling}
Let $k$ denote the number of vertical boundaries of $A$ and $T_v= \sum_{i=1}^{k}t(B_i)$. Then array $A$ can be $\alpha L$- tiled with $T_v+1$ tiles.
\end{restatable}

\dowod Suppose first that for each $1\leq i\leq k$ it holds that $B_i=B'_i$. Then by Definition \ref{def:boundary}, each subarray $A[b_i .. b_{i+1}-1]$ can be tiled horizontally with $t(B_i)$ tiles and the subarray $A[1..b_1-1]$ can be covered by $1$ tile. Therefore we indeed use $T_v+1$ tiles.

For the general case, let $i= \min\{k: B_k \neq B'_k\}$. It means that the subarray $A[b_{i-1} .. b_i]$ can be tiled horizontally with $t(B_i)$ tiles. By Definition $\ref{def:boundary}$ for each  $j\leq i-2$ the subarray $A[b_j .. b_{j+1}-1]$ can be tiled horizontally with $t(B_j)$ tiles and the subarray $A[1..b_1-1]$ can be covered by a single tile. This way the number of used tiles amounts to $\sum_{j=1}^{i-2}t(B_j) + t(B_i) +1 \leq  \sum_{j=1}^{i}t(B_j)$. We continue in the same manner with the subarray $A[b'_i..n]$. \koniec

Analogously, we define a horizontal sequence of boundaries of $A$, i.e., a vertical sequence of boundaries of $A^T$.

Throughout the paper, $B_1, B_2, \ldots, B_k$  and $C_1, \ldots, C_l$ denote, respectively, the vertical and horizontal sequence of boundaries of $A$.
Let $T_v= \sum_{i=1}^{k}t(B_i)$ and $T_h=\sum_{i=1}^{l}t(C_i)$ and let $T=\min\{T_v, T_h\}$.

\begin{fact} \label{F1}
Array $A$ can be $\alpha L$-tiled with $T+1$ tiles. 
\end{fact}

Since we can always $\alpha L$-partition $A$ into $T+1$ tiles, to prove that there exists an $\alpha$-approximation algorithm for the binary 
\RTILE problem, it suffices to show that $T+1$ is an allowed number of tiles, i.e., that $T+1 \leq p$. To do so, it is enough to prove that
it always holds that $w(A) > TL$. This is because since $w(A) \leq pL$ and  $T$ and $p$ are integers, the inequality $w(A) > TL$ implies $T+1 \leq p$.

We state this observation as:
\begin{fact} \label{F2}
Let $\alpha$ be such that for any $A$ it holds that $w(A)>TL$. Then $p \geq T+1$ and there exists an $\alpha$-approximation algorithm for the binary 
\RTILE problem.
\end{fact}

Let us first note that it is easy to prove that $w(A) > \frac{TL}{2}$.
\begin{restatable}{lemma}{lemmatwo}\label{Thalf}
The weight of $A$ satisfies $w(A) > \frac{TL}{2}$. Hence, $p>\frac{T}{2}$.
\end{restatable}

\dowod
We begin by proving that it is always possible to partition the array $A$ vertically into disjoint subarrays $A_1, A_2, \ldots, A_k$ where each subarray except the first ($A_1$) will contain one of the following boundary types:
\begin{enumerate}
    \item A single boundary of type $j$, where $j$ is greater than one.
    \item Two boundaries of type $1$.
    \item A boundary of type $j$ followed by a boundary of type $1$, where $j$ is greater than one.
\end{enumerate}

The first subarray $A_1$ may have any of the types of boundaries mentioned above or a single boundary of type $1$. We will now describe how to construct these subarrays.

Let $B_1, B_2, \ldots, B_l$ be a sequence of vertical boundaries of $A$. We construct the subarrays $A_1, \ldots, A_k$ iteratively. If $t(B_l) > 1$, then we define $A[b_l .. n]$ as the last vertical subarray, otherwise, the last subarray is represented by $A[b_{l-1} .. n]$. We then repeat this process on the remaining array $A[1 ... b_{x-1}]$, where $x \in \{l-1, l\}$ based on our choice of the last vertical subarray. We continue until we cannot construct a vertical subarray with one of the sets of boundaries mentioned in points $1 - 3$. 

In this case, the remaining subarray either has no boundary or has a boundary of type $1$. If it has no boundary, we merge it with the vertical subarray containing the first boundary. If it has a boundary of type $1$, we define it as the first vertical subarray and call it $A_1$.


Suppose $T_i$ represents the sum of the types of boundaries that are located within the subarray $A_i$. Our goal is to prove that $w(A_i) \geq \frac{1}{2}T_iL$. If  $A_i$ contains a boundary $B_r$ of type $j > 1$ then,
\begin{align*}
    & w(A_i) \geq w(B_r) \geq (j-1).\frac{3}{2}L 
     \geq \frac{1}{3}(j+1).\frac{3}{2}L \\ &\geq \frac{1}{2}t(B_r)L = \frac{1}{2}T_iL
\end{align*}

If $A_i$ contains two boundaries $B_r$ and $B_{r+1}$ of type $1$ then,
\begin{align*}
&w(A_i) \geq w(A[b_r \ldots b_{r+1}]) \geq \frac{3}{2}L \\&\geq \frac{1}{2}(t(B_r) + t(B_{r+1}))\frac{3}{2}L > \frac{1}{2}T_i.L
\end{align*}

If $B_r$ is a boundary of type $j>1$ and $B_{r+1}$ is a boundary of type $1$ in $A_i$ then, 
\begin{align*}
    & w(A_1) \geq w(B_r) \geq (j-1).\frac{3}{2}L \geq \frac{1}{3}(j+1).\frac{3}{2}L \\& \geq \frac{1}{2}(t(B_r) + t(B_{r+1}))L = \frac{1}{2}T_i.L
\end{align*}

If the $A_1$ contains one of the above sets of boundaries then $w(A_1) \geq \frac{T_1.L}{2}$. Otherwise, $A_1$ contains a single boundary of type $1$. Then, 

$$w(A_1) \geq \frac{3}{2}L > \frac{1}{2}t(B_1)L = \frac{1}{2}T_1.L$$

In conclusion we have proved that  for each $A_i$, $w(A_i) \geq \frac{T_i.L}{2}$. Therefore $w(A) \geq \frac{TL}{2}$.
\koniec

We now present a lemma that establishes conditions under which a subarray $A'$ of $A$ can be partitioned into horizontal tiles.

\begin{restatable}{lemma}{lemmathree}\label{3col}

Let $A'=A[i_1\ldots i_2]$ be a subarray of $A$ and $k$ a natural number greater or equal $2$.

Suppose that $\frac{w(A[i_1])+w(A[i_2])}{k} + w(A[i_1+1 \ldots i_2-1]) \leq \alpha L$. Then $A'$ can be partitioned into $k$ horizontal tiles of weight at most: (i) $\alpha L +1$ if $k=2$, (ii) $\alpha L +2$ if $k>2$.
\end{restatable}

\dowod We  divide the number equal to the sum of the weights of two     columns $A[i_1]$ and $A[i_2]$ (the columns may  be unconnected) by $k$. We check where the division lines fall with respect to the subarray. Often they may occur in the middle of a row (consisting of two array elements) and we have to move  the division so that the whole row is included or the whole row is excluded. If $k=2$, then we choose one of the two options - moving the division upwards or downwards, hence, in the worst case we may have to increase the weight of one tile by $1$.  For $k>2$, we may have to shift the division by almost the whole row and thus increase the weight of some tiles by $2$.   Next we extend this partition to include the subarray $A''=A[i_1+1 \ldots i_2-1]$ - we do not change the partition of the two-column subarray, but simply follow the partition lines. In the worst case the 
whole weight of $A''$ will fall into only one tile - yielding a tile of weight $\frac{w(A[i_1])+w(A[i_2])}{k} + w(A[i_1+1 \ldots i_2-1])$.
\koniec

\section{\texorpdfstring{A $\frac{3}{2}$-approximation when $w(A) \gg p^2$}{A 3/2-approximation when w(A) is much greater than p squared}} \label{easyalgorithm}

In this section we deal with arrays such that $\frac{p^2}{w(A)}$ is close to $0$, which means that the total weight of any $p^2$ elements
of $A$ is negligible. We are going to show that under this assumption, for $\alpha=\frac{3}{2}$, the weight of $A$ satisfies $w(A)>TL$.
Hence, by Fact \ref{F2} we get that for this type of arrays there exists a $\frac{3}{2}$-approximation for the binary \RTILE problem.
For the general case the proof that $\alpha=\frac{3}{2}+ \frac{p^2}{w(A)}$, the weight of $A$ satisfies $w(A)>TL$, presented in the next section will be only a slight modification of the one shown here.

\begin{convention}
Throughout this section whenever we speak about \textbf{\emph{tiling}} and \textbf{\emph{partitioning}}, we respectively mean ``$\frac{3}{2}L$-partitioning'' and ``tiling using tiles of weight at most $\frac{3}{2}L$''.
\end{convention}

\textbf{Remark:} The total number of cells at the intersection of the horizontal and vertical boundaries is  $O(T^2)$. By Lemma \ref{Thalf} we have that $O(T^2)=O(p^2)$. Therefore by the assumption of this section, it implies that the total weight of the cells in the intersections of the boundaries is negligible with respect to the total weight of the array.

\begin{observation}\cite{paluch2004approximation}\label{obs:thesis}
Assume we have two complexes: one as in Figure \ref{fig:observation1} and the other with the variables related to the variables of the first one as follows: \( x'_1 = x_1 \), \( x'_4 = x_4 \), \( x'_3 = x_3 \), \( x'_5 = x'_6 = 0 \) and \( x'_2 = x_2 + \max\{x_5, x_6\} \). Then the weight of the second complex is not bigger than the weight of the first one while the inequalities describing the first complex remain true for the second.
\end{observation}
\begin{figure}
    \centering
    \begin{tikzpicture}
    \draw[thick] (0, 0) rectangle (4, 5);
    
    \draw[thick] (2, 0) -- (2, 5);
    \draw[thick] (0, 2) -- (4, 2);
    \draw[thick] (0, 3) -- (4, 3);
    \node at (1, 1) {$x_6$};
    \node at (3, 1) {$x_4$};
    \node at (1, 2.5) {$x_2$};
    \node at (3, 2.5) {$x_1$};
    \node at (1, 4) {$x_5$};
    \node at (3, 4) {$x_3$};
    \end{tikzpicture}
    \caption{An array with one horizontal boundary containing the subarrays $x_2$ and $x_1$ and one vertical boundary containing the subarrays $x_3$, $x_1$ and $x_4$}
    \label{fig:observation1}
\end{figure}

Given an array $A$ we build a linear program, with the help of which we will be able to relate the total weight of the array to the sum of types of boundaries $T$, i.e., we will show that $w(A)>TL$.

Using Observation \ref{obs:thesis}, we can assume that the whole weight of the array $A$ is contained in the boundaries, i.e., each element of $A$ that does not belong to any boundary has value $0$.
Each vertical boundary $B_i$ is crossed by $l$ horizontal boundaries and thus cut into $l+1$ parts. We assign a variable $x_{j,i}$ to each part, i.e., the $j$th part of $B_i$ consists of  elements $A[c_{j-1}+1, b_i], A[c_{j-1}+2, b_i], \ldots, A[c_j-1,b_i]$ and $x_{j,i}$ denotes the sum of the weights of these elements. Similarly, each horizontal boundary $C_i$ is crossed by $k$ vertical boundaries and thus cut into $k+1$ parts. We assign a variable $z_{i,j}$ to each such part. The value of each variable $x_{j,i}$ or $z_{i,j}$ denotes the weight of the corresponding part of the boundary.

In the linear program, we minimize the sum of non-negative variables $x_{j,i}$ and $z_{i,j}$ subject to a set of constraints associated with the boundaries. 
For each vertical boundary $B_i$ we will have either one or two constraints of the following form:

\begin{enumerate}
\item If $t(B_i)>1$, then  we add the constraint $\frac{1}{t(B_i)-1}\sum_{j=1}^{l+1} x_{j,i} \geq \frac{3}{2}L$, which simply describes the total weight
of $B_i$.
\item \begin{enumerate}
\item $B'_{i-1} \neq B_{i-1}$. 
\begin{enumerate}
\item $t(B_i)=1$. The added constraint is $\sum_{j=1}^{l} z_{j,i}+ \sum_{j=1}^{l+1} x_{j,i} >\frac{3}{2}L.$
\item $t(B_i)>1$ and $B'_i=B_i$ (which means that $A[b'_{i-1}..b_{i}]$ cannot be tiled horizontally with $t(B_i)$ tiles).
By Lemma \ref{3col} we are justified to add the constraint $\sum_{j=1}^{l} z_{j,i}+ \frac{1}{t(B_i)}\sum_{j=1}^{l+1} x_{j,i} >\frac{3}{2}L.$
\item $t(B_i)>1$ and $B'_i\neq B_i$. In this case we do not add any constraint.
\end{enumerate}

\item $B'_{i-1}=B_{i-1}$.

\begin{enumerate}
\item $B'_i \neq B_i$. The added constraint is $\sum_{j=1}^{l} z_{j,i}+ \frac{1}{t(B_{i-1})}(\sum_{j=1}^{l+1} x_{j,i} + \sum_{j=1}^{l+1} x_{j,i-1}) >\frac{3}{2}L$.

\item $B'_i = B_i$. Let $T_i=\max\{t(B_{i-1}), t(B_i)\}$. The constraint we add is  $\sum_{j=1}^{l} z_{j,i}+ \frac{1}{T_i}(\sum_{j=1}^{l+1} x_{j,i} + \sum_{j=1}^{l+1} x_{j,i-1}) >\frac{3}{2}L$. The constraint is a consequence of Lemma \ref{3col}.
\end{enumerate}
\end{enumerate}
\end{enumerate}

Thus, each $B_i$ defines either one or two constraints. 
Analogously, each horizontal variable $C_j$ also defines one or two constraints. 
The linear program dual to the one we have just described has dual variables $y'_i, y_i$. For each $B_i$ with $t(B_i)>1$ 
let $y'_i$ denote the dual variable corresponding to the constraint $\frac{1}{t(B_i)-1}\sum_{j=1}^{l+1} x_{j,i} >\frac{3}{2}L$. The other type of a constraint (if it exists) defined by $B_i$ is represented by $y_i$. The dual variables corresponding to horizontal boundaries are $w_i, w'_i$.\\

\begin{example}\label{example4}
In this example array $A$ has one vertical boundary $B_1$ of type $1$ and one horizontal boundary $C_1$ of type $1$.

\begin{figure}[h]
\centering{\includegraphics[scale=0.8]{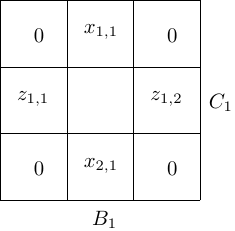}}
\caption{An array with one vertical and one horizontal boundary
} \label{ex5}
\end{figure}

The linear program for $A$ looks as follows. In brackets we give the dual variables corresponding to respective inequalities.

$\begin{array}{lrr }
\mbox{minimize} & x_{1,1}+x_{2,1}+z_{1,1}+z_{1,2} &\\
\mbox{subject to} & x_{1,1}+x_{2,1}+z_{1,1} > \frac{3}{2}L  & (y_1)\\
& x_{1,1}+z_{1,1}+z_{1,2} > \frac{3}{2}L & (w_1) 

\end{array}$ \\

\vspace{0.5cm}
\end{example}

\begin{example}\label{example5}
In this example array $A$ has two vertical boundaries and two horizontal ones, each of the four boundaries is of type $1$. The array is depicted in Figure \ref{ex6}. The linear program for $A$ looks as follows. In brackets we give the dual variables corresponding to respective inequalities.

$\begin{array}{lrr}
\mbox{minimize} & \sum_{i=1}^2 \sum_{j=1}^3 x_{j,i} + \sum_{i=1}^2 \sum_{j=1}^3 z_{i,j} &\\
\mbox{subject to} & \sum_{j=1}^3 x_{j,1} + \sum_{i=1}^2 z_{i,1} > \frac{3}{2}L & (y_1)\\
& \sum_{i=1}^2 \sum_{j=1}^3  x_{j,i} + \sum_{i=1}^2 z_{i,2}> \frac{3}{2}L & (y_2)\\
& \sum_{j=1}^3 z_{1,j} + \sum_{i=1}^2 x_{1,i} > \frac{3}{2}L & (w_1) \\
& \sum_{i=1}^2 \sum_{j=1}^3 z_{i,j} + \sum_{i=1}^2 x_{2,i} > \frac{3}{2}L & (w_2)

\end{array}$\\
\vspace{0.5cm}
\end{example}

\begin{example}\label{example6}
In this example array $A$ has two vertical boundaries $B_1, B_2$ and two horizontal ones $C_1, C_2$. Their types are the following:
$t(B_1)=t(C_2)=2$ and $t(B_2)=t(C_1)=1$. Also $B'_1 \neq B_1$ and $C'_2=C_2$.\\
\begin{figure}
\centering{\includegraphics[scale=0.8]{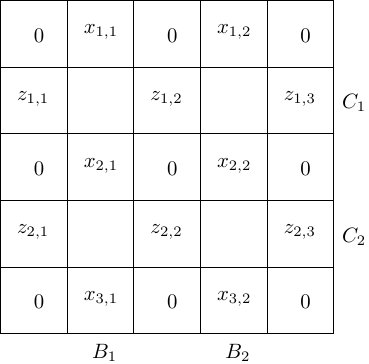}}
\caption{An array with two vertical and two horizontal boundaries
} \label{ex6}
\end{figure}
$\begin{array}{lrr}
\mbox{minimize} & \sum_{i=1}^2 \sum_{j=1}^3 x_{j,i} + \sum_{i=1}^2 \sum_{j=1}^3 z_{i,j}&\\
\mbox{subject to} \\
& \frac{1}{2}\sum_{j=1}^3 x_{j,1} + \sum_{i=1}^2 z_{i,1} > \frac{3}{2}L & (y_1)\\
& \sum_{j=1}^3 x_{j,1} >\frac{3}{2}L & (y'_1) \\
& \sum_{i=1}^2 z_{i,2} + \sum_{j=1}^3 x_{j,2}> \frac{3}{2}L & (y_2)\\
& \sum_{j=1}^3 z_{1,j} + \sum_{i=1}^2 x_{1,i} > \frac{3}{2}L & (w_1)\\
& \frac{1}{2} \sum_{i=1}^2 \sum_{j=1}^3  z_{i,j} + \sum_{i=1}^2 x_{2,i} > \frac{3}{2}L & (w_2)\\

& \sum_{j=1}^3 z_{2,j} > \frac{3}{2}L & (w'_2)\\

\end{array}$\\
\end{example}

Let us now build dual linear program for the primal linear program of the  Example \ref{example6}. The dual linear program has the  form:\\

$\begin{array}{lrr}
\mbox{maximize} & \frac{3}{2}L(y_1 +y_2 +w_1+w_2) &\\
\mbox{subject to} & y'_1+\frac{1}{2}y_1  + w_1 \leq 1 & (x_{1,1})\\
&  y_2 + w_1 \leq 1  & (x_{1,2})\\
& y'_1+\frac{1}{2}y_1  + w_2 \leq 1 & (x_{2,1})\\
& y_2 + w_2 \leq 1 & (x_{2,2}) \\
& y_1 + w_1 + \frac{1}{2} w_2 \leq 1 & (z_{1,1})\\
& y_2 + w_1 + \frac{1}{2} w_2 \leq 1 & (z_{1,2})\\
& y_1 + w'_2 + \frac{1}{2} w_2 \leq 1 & (z_{2,1})\\
& y_2 + w'_2 + \frac{1}{2} w_2 \leq 1 & (z_{2,2})\\
\end{array}$\\

\begin{algorithm}[t] 
\begin{footnotesize}
\caption{}
\label{main}
\begin{algorithmic}[1]
\State $A \leftarrow [1 \ldots n, 1 \ldots n]$  a two-dimensional array
\State Construct the horizontal and vertical boundaries and their shadows using Definition \ref{def:boundary}.

\State $B \leftarrow \{B_1, B_2, \ldots B_k\}$ (the vertical boundaries, where each $B_i=A[b_i]$)
\State $B' \leftarrow \{B'_1, B'_2, \ldots B'_k\}$ (the shadows of $B_i$s, where each $B'_i=A[b'_i]$)
\State $t(B) \leftarrow \{t(B_1), t(B_2), \ldots t(B_k)\}$ (the types of the vertical boundaries)

\State $C \leftarrow \{C_1, C_2, \ldots C_l\}$ (the horizontal boundaries, where each $C_i=A[c_i]$)
\State $C' \leftarrow \{C'_1, C'_2, \ldots C'_l\}$ (the horizontal boundaries, where each $C'_i=A[c'_i]$)
\State $t(C) \leftarrow \{t(C_1), t(C_2), \ldots t(C_l)\}$ (the types of the horizontal boundaries)

\State $T_v \leftarrow \sum_{i=1}^k {t(B_i)}$
\State $T_h \leftarrow \sum_{i=1}^l {t(C_i)}$

\If {$T_v \leq T_h$} 
\State use the vertical boundaries $B$ as described below from line $15$
\Else 
\State use the horizontal boundaries $C$ instead of the vertical ones
\EndIf	
\If {${B_k = B'_k}$}
    \State partition $A[b_k \ldots n]$ horizontally into $t(B_k)$ tiles
\Else 
    \State cover $A[b_k+1 \ldots n]$ with one tile
\EndIf
\For {$i = k-1 \ldots, 1$}
	
	\If {${B_i = B'_i}$}
    	\If {${B_{i+1} = B'_{i+1}}$}
		    \State tile $A[b_i \ldots b_{i+1}-1]$ horiz. into $t(B_i)$ tiles (by point \ref{3a} of Definition \ref{def:boundary})
		\Else 
		    \State partition $A[b_i \ldots b_{i+1}]$ horiz. into $t(B_{i+1})$ tiles (by point \ref{4b} of Definition \ref{def:boundary})
		\EndIf
    \Else $(B_{i+1} \neq B'_{i+1})$
    	\If {${B_{i+1} = B'_{i+1}}$}
    	    \State cover $A[b_i+1 \ldots b_{i+1}-1]$ horiz. with one tile (by point \ref{3b} of Definition \ref{def:boundary})
        \Else
            \State partition $A[b_i+1 \ldots b_{i+1}]$ horiz. into $t(B_{i+1})$ tiles (by point \ref{4b} of Definition \ref{def:boundary})
        \EndIf
    \EndIf
    \EndFor
\If {${B_1 = B'_1}$}
    \State cover $A[1 \ldots b_1-1]$ with one tile 
\Else 
    \State partition $A[1 \ldots b_1]$ horiz. into $t(B_1)$ tiles 
\EndIf\end{algorithmic}
\end{footnotesize}
\end{algorithm}

To figure out the form of constraints constituting the dual program in general,
let us consider a variable $x_{j,i}$. Notice that it occurs in at most one constraint defined by a horizontal boundary. It can possibly be contained only in the constraint defined by $C_j$ represented by $w_j$, where its coefficient is $1$. 
If $t(B_i)=1$, then we do not have $y'_i$ and $x_{j,i}$  occurs in the constraint represented by $y_i$ and possibly in the constraint represented by
$y_{i+1}$. Thus the inequality in the dual program corresponding to $x_{j,i}$ has the form $\alpha_{j,i}y_i+\alpha_{j,i+1}y_{i+1}+\beta_{j,i}w_j \leq 1$, where each of the coefficients belongs to $[0,1]$. 

If $t(B_i)>1$, then we do have $y'_i$ and $x_{j,i}$  occurs in this constraint with the coefficient $\frac{1}{t(B_i)-1}$. If $B_i \neq B'_i$, then $x_{j,i}$ does not occur in any other constraints and the inequality in the dual program has the form $ \frac{1}{t(B_i)-1}y'_i+\beta_{j,i}w_j \leq 1$, where $\beta_{j,i} \in [0,1]$. Otherwise, $x_{j,i}$ may also belong to the constraints represented by $y_i$ and $y_{i+1}$. In each one of them it occurs with the coefficient equal to at most $\frac{1}{t(B_i)}$. Therefore the inequality has the form $ \frac{1}{t(B_i)-1}y'_i+ \alpha_{j,i}y_i+\alpha_{j,i+1}y_{i+1}+\beta_{j,i}w_j \leq 1$, where each of the coefficients $\alpha_{j,i+1},\alpha_{j,i}$ belongs to $[0,\frac{1}{t(B_i)}]$.


We are  ready to lower bound the weight of the array $A$ with the aid of its boundaries and their shadows.

\begin{lemma} \label{weight}
 $w(A)>TL$. 
\end{lemma}
\dowod Since the value of any cost function of the dual linear program described above is a lower bound on the minimal value of the cost function of the primal linear program,
it suffices to find a feasible assignment of the dual variables such that the cost function will be have value greater than $TL$.

\begin{claim}
We can satisfy all constraints of the dual program by assigning the following values to the dual variables. 
Each $y_i$ and each $w_j$ is assigned $\frac{1}{3}$. If $B_i$ defines only one constraint $y'_i$, then we assign $\frac{t(B_i)}{3}$ to $y'_i$.
Otherwise $y'_i$ is assigned  $\frac{t(B_i)-1}{3}$. 
\end{claim}

The claim follows from the fact that the inequality $\frac{1}{T-1} \cdot \frac{T}{3}+\frac{1}{3} \leq 1$ is satisfied by each $T\geq 2$
and that the inequality $\frac{T-1}{3(T-1)} + \frac{2}{3T}+\frac{1}{3} \leq 1$ is also satisfied by each $T\geq 2$.

This means that the total value contributed by the dual variables $y_i, y'_i$ (corresponding to constraints defined by boundary $B_i$) is at least
$\frac{t(B_i)}{3}$.

Thus $w(A) > \frac{3}{2}L (\frac{T_v}{3} + \frac{T_h}{3}) \geq TL$.
\koniec

We show a method for finding a tiling of $A$ with at most $T + 1$ tiles. By Facts \ref{nrtiles}, \ref{F1}, Lemmas \ref{tiling} and \ref{weight}, we proved that $T+1 \leq p$, when the approximation factor is $\frac{3}{2}$. In other words, we obtain a $\frac{3}{2}$-approximation algorithm for binary \RTILE.

\begin{theorem}
For any array $A$ satisfying $\frac{p^2}{w(A)} \approx 0$, there exists a linear time $\frac{3}{2}$-approximation algorithm for binary \RTILE.
\end{theorem}

\section{\texorpdfstring{A $(\frac{3}{2}+\beta)$-approximation}{A (3/2 + beta)-approximation}} \label{hardalgoriithm}

In this section we examine the general case, arrays such that $\frac{p^2}{w(A)}$ is not negligible. 
We will aim for a $(\frac{3}{2}+\beta)$-approximation. When $\beta <\frac{1}{2}$, the approximation ratio of our algorithm is better than $2$.
Throughout the section, whenever we refer to tiling and partitioning, we mean  $(\frac{3}{2}+\beta)L$-partitioning and tiling using tiles of weight at most $(\frac{3}{2}+\beta)L$.

We define a sequence of boundaries and shadows analogously as in the previous section, but  with respect to  $(\frac{3}{2}+\beta)L$, i.e., we replace each occurrence of ``$\frac{3}{2}L$''  with ``$(\frac{3}{2}+\beta)L$'' and modify the meaning of tiling and partitioning accordingly, i.e., to $(\frac{3}{2}+\beta)L$-partitioning.

We want to prove an analogue of Lemma \ref{weight}. To this end we will consider an analogous linear program, in which we have all the variables occurring in the previous section and additionally we have a variable $s_{i,j}$ for each pair $(B_i, C_j)$, which denotes the element of $A$ at the intersection of the vertical boundary $B_i$ and the horizontal  boundary $C_j$. The function we minimize is $\sum_{i=1}^k x_{j,i} +  \sum_{j=1}^l z{i,j} + \sum_{i=1}^k \sum_{j=1}^l s_{i,j}$.
For each variable $s_{i,j}$ we have an additional constraint: $-s_{i,j} \geq -1$. The variable $s_{i,j}$ is also included in all those constraints which refer to the part
of $A$ covering the intersection of $B_i$ with $C_j$.

For instance, the linear program for the array from Example \ref{example4} is modified as follows:

$\begin{array}{lrr }
\mbox{minimize} & x_{1,1}+x_{2,1}+z_{1,1}+z_{1,2} +s_{1,1} &\\
\mbox{subject to} & x_{1,1}+x_{2,1}+z_{1,1} +s_{1,1}> (\frac{3}{2}+\beta)L  & (y_1)\\
& x_{1,1}+z_{1,1}+z_{1,2} + s_{1,1}> (\frac{3}{2}+\beta)L & (w_1)  \\
& -s_{1,1} \geq -1 & (t_{1,1}). 

\end{array}$\\

The linear program for the array from Example \ref{example5} in the new scenario looks as follows:\\

${\fontsize{8}{9}\selectfont 
\begin{array}{lrr}

\mbox{min}&  \sum_{i=1}^2 \sum_{j=1}^3 x_{j,i} + \sum_{i=1}^2 \sum_{j=1}^3 z_{i,j} +\sum_{i=1}^2 \sum_{j=1}^2 s_{i,j} &\\
\mbox{s.t.}&  \sum_{j=1}^3 x_{j,1} + \sum_{i=1}^2 z_{i,1} + \sum_{j=1}^2 s_{1,j} > (\frac{3}{2}+\beta)L & (y_1)\\
& \sum_{i=1}^2 \sum_{j=1}^3 x_{j,i} + \sum_{i=1}^2 z_{i,2} + \sum_{i=1}^2 \sum_{j=1}^2 s_{i,j} > (\frac{3}{2}+\beta)L  & (y_2)\\
& \sum_{j=1}^3 z_{1,j} + \sum_{i=1}^2 x_{1,i} + \sum_{i=1}^2 s_{i,1} > (\frac{3}{2}+\beta)L  & (w_1) \\
& \sum_{i=1}^2 \sum_{j=1}^3 z_{i,j} + \sum_{i=1}^2 x_{2,i} + \sum_{i=1}^2 \sum_{j=1}^2 s_{i,j}> (\frac{3}{2}+\beta)L  & (w_2)\\
& -s_{i,j} \geq - 1  (t_{i,j}) \mbox{, \ for each\ } 1 \leq i,j \leq 2.&

\end{array}}$\\

Correspondingly, in the dual program we maximize $(\frac{3}{2}+\beta)L(\sum{y_i}+\sum{w_j}) - \sum{t_{i,j}}$ and we have an additional constraint for each primal variable $s_{i,j}$.

The dual linear program for the array from Example \ref{example5} contains the following additional inequalities.\\

$\begin{array}{lr}
 y_1 + y_2 + w_1 +w_2 - t_{1,1} \leq 1 & (s_{1,1})\\

 y_1 + y_2 + w_2 -t_{1,2} \leq 1 & (s_{1,2})\\

 y_2 + w_1 + w_2 -t_{2,1} \leq 1 & (s_{2,1})\\
 y_2  + w_2 -t_{2,2} \leq 1 & (s_{2,2}).
\end{array}$\\

We can see that if we want to assign $\frac{1}{3}$ to each variable $y_i, w_j$, then we sometimes also have to assign $\frac 13$ to variables $t_{i,j}$ to ensure the feasibility  - compare the first inequality in the set of additional inequalities above. We can notice that
we have to assign $\frac 13$ to $t_{i,j}$ only if both  $i<k$ and $j<l$, i.e. when $s_{i,j}$ does not belong to $B_k$ or $C_l$.

\begin{restatable}{lemma}{lemmafive} \label{weight1}
 For $\beta= \frac{p^2}{w(A)}$, it holds that $w(A)>TL$. 
\end{restatable}

\dowod We can satisfy all constraints of the dual program by assigning the following values to the dual variables. 
Each $y_i$ and each $w_j$ is assigned $\frac{1}{3}$. If $B_i$ defines only one constraint $y'_i$, then we assign $\frac{t(B_i)}{3}$ to $y'_i$.
Otherwise $y'_i$ is assigned  $\frac{t(B_i)-1}{3}$. Also, each $t_{i,j}$ such that $i<k$ and $j<l$ is assigned $\frac{1}{3}$.

Some of the constraints in the primal program have  value $(\frac{3}{2}+\beta)L -2$ on the right hand side. Such constraints correspond to some borders of type greater than $2$, when we use Lemma \ref{3col}. Let us analyze such cases in more detail. Assume that  for a boundary $B_i$ of type $k>2$  we indeed use Lemma \ref{3col}. Then the primal linear program contains  a constraint with value $(\frac{3}{2}+\beta)L -2$ on the right hand side. This constraint corresponds to the dual variable $y_i$. We notice that  $B_i$ also defines a constraint of type $1$, which has  $(\frac{3}{2}+\beta)L$ on the right hand side and corresponds to the dual variable $y'_i$. Hence each such boundary contributes at least $\frac{t(B_i)-1}{3}(\frac{3}{2}+\beta)L+ \frac{1}{3}(\frac{3}{2}+\beta)L -2)\geq (\frac{(\frac{3}{2}+\beta)L -\frac{2}{3}t(B_i)}{3})$  to the cost function of the dual linear program.

Similarly, some of the constraints in the primal program have  value $(\frac{3}{2}+\beta)L -1$ on the right hand side. Such constraints correspond to some borders of type equal to $2$, when we use Lemma \ref{3col}. Each such boundary contributes at least $(\frac{(\frac{3}{2}+\beta)L -\frac{1}{2}t(B_i)}{3})$  to the cost function of the dual linear program.

Thus the value of the cost function of the dual linear program  is lower bounded by  $ ((\frac{3}{2}+\beta)L -\frac{2}{3}) (\frac{T_v}{3} + \frac{T_h}{3}) - \frac{(T_h-1)(T_v-1)}{3}
\geq TL +p\frac{T_v+T_h}{3} -\frac{2}{3}(T_v+T_h) - \frac{(T_h-1)(T_v-1)}{3}$. Since $L \geq \frac{w(A)}{p}$, we get that $w(A) \geq TL+ \frac{p(T_v+T_h)}{3}$ +$\frac{T_v+T_h}{9}-\frac{T_vT_h}{3} -\frac{1}{3}$. Because $p \geq T$, we obtain that 
$\frac{p(T_v+T_h)}{3}$ +$\frac{T_v+T_h}{9}-\frac{T_vT_h}{3} -\frac{1}{3} \geq \frac{T^2}{3}+\frac{T_v+T_h}{9} -\frac{1}{3}>0.$ 

Therefore, $w(A)>TL$. \koniec

\begin{theorem}
 There exists a $(\frac{3}{2}+ \frac{p^2}{w(A)})$-approximation algorithm for binary \RTILE that has a linear ($O(n^2)$) running time.
\end{theorem}

%% file: 4.Tightness.tex
\section{Tightness of approximation} \label{tightness}
\begin{figure}\label{fig:tightness}
\centering
\input{cross}
\caption{The empty squares denote a value of $0$, while the ones are colored black. $(a)$ On tiling this array with $3$ tiles, one tile will always contain $5$ ones, giving an approximation factor of $\frac{5}{3}$. $(b)$ $4$-crosses placed in an array for proving an approximation lower bound of $\apxfac$.}
\label{fig:1p5}
\end{figure}
In this section, we show that the approximation ratio for the \RTILE problem is tight with respect to the only known lower bound. Percisely, we prove the following theorem.
\begin{restatable}{theorem}{theoremone}\label{tightnessth}
Let $p=2k$, for some $k\in \mathbb{N}$. Then, there exists a binary array $A_k$ such that the maximum weight of a tile in any tiling of $A_k$ into $p$ tiles has weight at least $\apxfac \cdot \frac{w(A_k)}{p}+{1}$.
\end{restatable}

We define an $L$-{\em cross} to consist of $2L+1$ ones, it is obtained by taking a $(L+1) \times (L+1)$ array, and filling  the $(\frac{L}{2}+1)^{th}$ row as well as the $(\frac{L}{2}+1)^{th}$ column with ones, finally the rest of the entries are filled with zeros.
The coordinate $(\frac{L}{2}+1,\frac{L}{2}+1)$ is referred to as the center of the $L$-cross defined above.
An $L$-cross centered at $(x,y)$ is obtained by translating the center of an $L$-cross to the coordinate $(x,y)$. Note that an $L$-cross consists of four contiguous segments of ones, referred to as {\em arms}, each containing $\frac{L}{2}$ ones. We define $A_k$ as shown in Figure~3.

\dowod
Suppose that $p$ is even, therefore $p=2k$, for some $k\in \mathbb{N}$; our input binary array $A_k$ is obtained as follows. We place $k$ many $L$-crosses centered at $(j\cdot(L+1)-\frac{L}{2}, j\cdot(L+1)-\frac{L}{2})$, for each $1\leq j \leq k$, the rest of the entries of $A_k$ are zero.
Note that the $L$-crosses are placed diagonally in a non-overlapping manner, and $\frac{w(A)}{p} = L$.  The array for $L=4$ is illustrated in Figure~\ref{fig:1p5}$(a)$.

If $p=2$, then it is obvious that one tile will have to contain $3$ arms of the cross and thus have weight $\frac{3}{2}L+1$.
We will prove that for every $k \in \mathbb{N}$, one of the tiles will have weight at least $\frac{3}{2}L+1$.

Suppose that for $k$ crosses and $2k$ tiles the thesis holds by induction. We will now prove it for $k+1$ crosses and $2k+2$ tiles.
Let $T_1$ be the tile that contains the cell $A_{k+1}[1,1]$. If this tile has weight smaller than $\frac{3}{2}L$, then we have the following two cases:

\begin{enumerate}
\item If $T_1$ does not contain the center of the lower left cross $A_{k+1}[\frac{L}{2}+1, \frac{L}{2}+1]$, then $T_1$ is formed either by the first $\frac{L}{2}$ columns or the first $\frac{L}{2}$ rows. Due to symmetry, it is enough to consider the case when $T_1$ is formed by the first $\frac{L}{2}$ columns. Let $T_2$ be the tile that contains $A_{k+1}[\frac{L}{2}+1, 1]$. If $T_2$ has weight smaller than $\frac{3}{2}L$, then its upper right corner $A_{k+1}[x,y]$  is such that either $x<L+1$ or $y<L+1$. Due to symmetry, it is enough to consider the case when $x<L+1$. In this case, if $y<n$, then we can extend $T_2$ so that $y=n$ without increasing the weight of $T_2$ as it would not intersect any new $L$-cross.

Thus, we are left with $2k$ tiles, and an array which has $A_k$ as a subarray, therefore by the induction hypothesis we get that the weight of maximum weight tile is at least $\frac{3}{2}L+1$.

\item If $T_1$  contains the center of the cross $A_{k+1}[\frac{L}{2}+1, \frac{L}{2}+1]$, then its right upper corner $A_{k+1}[x,y]$ is such that $x<L+1$ or  $y<L+1$.
Notice that we may assume that either one tile will be formed by subarray $A_{k+1}[x+1,1,n,y]$ or by subarray $A_{k+1}[1,y+1, x,n]$. This is because the tile that contains $A_{k+1}[x+1, y+1]$ cannot contain both $A_{k+1}[1, y+1]$ and $A_{k+1}[x+1, 1]$. Suppose w.l.o.g. that the tile $T_2$ that contains $A_{k+1}[x+1, 1]$ does not cover any cell of row $y+1$. We may then extend the upper right corner of $T_2$ till $A_{k+1}[n,y]$, without increasing the weight of any tile. We are left with $2k$ tiles, and the induction hypothesis gives us the result.
\end{enumerate}
In the case when $p$ is odd, the input binary array $A'_p$ is obtained from $A_{k+1}$ by deleting any rows and columns in it with index at least $k(L+1) + \frac{L}{2}+2$. This, in effect adds an extra half $L$-cross near the upper right corner of $A_k$. Clearly, for $p=3$ it is not possible to tile $A'_3$ with $3$ tiles such that the weight of maximum weight tile is less than $\frac{3}{2}L$. The rest of the proof follows from arguments similar to the case when $p$ is even.
\koniec

\textbf{Remark:} The approximation factor of our algorithm is $\frac{3}{2} + \frac{p^2}{w(A)}$ which is equal to $(\frac{3}{2} + \frac{p}{L})L$. Since $\frac{p}{L}$ is equal to $\frac{p^2}{w(A)}$, it means that the approximation of our algorithm is tight under the condition that $w(A) \gg p^2$.

%% file: cross.tex
\begin{tikzpicture}[scale = 0.65]
	\begin{pgfonlayer}{nodelayer}
		
	\begin{scope}[shift={(2.8,0)}]
	\draw[step=0.5] (0.5,0.5) grid (3,3);
		\node [] (0) at (1.75, -0.5) {$(a)$};
		\node [style=rect] (1) at (1.75, 0.75) {};
		\node [style=rect] (2) at (1.75, 1.25) {};
		\node [style=rect] (3) at (1.75, 1.75) {};
		\node [style=rect] (4) at (1.25, 1.75) {};
		\node [style=rect] (5) at (0.75, 1.75) {};
		\node [style=rect] (7) at (2.25, 1.75) {};
		\node [style=rect] (8) at (2.75, 1.75) {};
		\node [style=rect] (10) at (1.75, 2.25) {};
		\node [style=rect] (11) at (1.75, 2.75) {};
		\end{scope}
\begin{scope}[shift={(0,-10)}]

		\draw[step=0.5](0,0) grid (9,9);
		\node [] (27) at (4.5, -1) {$(b)$};
		\node [style=rect] (0) at (0.25, 1.25) {};
		\node [style=rect] (1) at (0.75, 1.25) {};
		\node [style=rect] (2) at (1.75, 1.25) {};
		\node [style=rect] (3) at (2.25, 1.25) {};
		\node [style=rect] (4) at (1.25, 0.75) {};
		\node [style=rect] (5) at (1.25, 0.25) {};
		\node [style=rect] (6) at (1.25, 1.75) {};
		\node [style=rect] (7) at (1.25, 2.25) {};
		\node [style=rect] (8) at (3.75, 3.25) {};
		\node [style=rect] (9) at (4.75, 3.75) {};
		\node [style=rect] (10) at (3.75, 4.75) {};
		\node [style=rect] (11) at (3.75, 4.25) {};
		\node [style=rect] (12) at (4.25, 3.75) {};
		\node [style=rect] (13) at (3.75, 2.75) {};
		\node [style=rect] (14) at (2.75, 3.75) {};
		\node [style=rect] (15) at (3.25, 3.75) {};
		\node [style=rect] (16) at (7.75, 7.25) {};
		\node [style=rect] (17) at (8.75, 7.75) {};
		\node [style=rect] (18) at (7.75, 8.75) {};
		\node [style=rect] (19) at (7.75, 8.25) {};
		\node [style=rect] (20) at (8.25, 7.75) {};
		\node [style=rect] (21) at (7.75, 6.75) {};
		\node [style=rect] (22) at (6.75, 7.75) {};
		\node [style=rect] (23) at (7.25, 7.75) {};
		\node [style=dott] (24) at (5.25, 5.25) {};
		\node [style=dott] (25) at (5.75, 5.75) {};
		\node [style=rect] (26) at (1.25, 1.25) {};
		\node [style=dott] (27) at (6.25, 6.25) {};
		\node [style=rect] (28) at (7.75, 7.75) {};
		\node [style=rect] (29) at (3.75, 3.75) {};

		\end{scope}	
	\end{pgfonlayer}
\end{tikzpicture}

%% file: 5.drtile.tex
\section{\DRTILE} \label{drtile}
In this section, we present an approximation algorithm for the \DRTILE problem. We have presented a $\apxfac+ \beta$-approximation algorithm for the \RTILE problem in Section $3$. Now we show how to reduce an instance of the \DRTILE problem to an instance of the \RTILE problem to achieve an approximation ratio for the \DRTILE problem. Before we proceed, let us recall the definition of the \DRTILE problem. 

\begin{itemize}
    \item \textbf{the \DRTILE problem}
\begin{itemize}
    \item \textbf{Input:} A two-dimensional array $A$ and a weight upper bound $w$.
    \item \textbf{Goal:} Partition $A$ into a minimum number of non-overlapping tiles, where the weight of each tile must not be larger than  $W$.
\end{itemize}
\end{itemize}

Let us consider an array $A$ with $w(A) = n$. 
Suppose $W$, provided as input, is the maximum allowed weight of any tile. Clearly, the minimal number of tiles we need to use to cover $A$ is $\ceil{\frac{n}{W}}$. Consequently, $\ceil{\frac{n}{W}}$ is a lower bound to the optimal solution of the \DRTILE problem. Our goal is to obtain a $\gamma$-approximation algorithm, where $\gamma$ depends on $W$. Therefore, the number of tiles we are allowed to use to cover $A$ with this approximation is $\gamma \times \ceil{\frac{n}{W}}$. 

We construct an instance of the the \RTILE problem as follows: as an input we have the same array $A$, and we are allowed to use at most $p = \gamma \times\ceil{{\frac{n}{W}}}$ tiles. Hence from Section \ref{section2}, the lower bound on the maximum weight of a tile is $\ceil{\frac{n}{\gamma \times\ceil{{\frac{n}{W}}}}}$. Hence the maximum weight of a tile with approximation factor of $\apxfac + \beta$ is,

\[
\ceil[\Big]
	{
	\frac	{n}
			{ \gamma \times \ceil{ {\frac{n}{W} } } }
	} \times (\apxfac+ \beta)\\
\leq \ceil[\Big]
	{
	\frac{n}{ \gamma \times \frac{n}{W}}
	} \times (\apxfac+ \beta)\\
= \ceil[\Big]{\frac{W}{\gamma}} \times (\apxfac+ \beta).
\]

For the solution returned by \RTILE to be a valid solution of \DRTILE, the value of $\ceil{\frac{W}{\gamma}} \times (\apxfac + \beta)$ must not exceed $W$. This allows us to derive a bound on the value of the approximation factor $\gamma$, we have, 

\begin{align*}
&\ceil{\frac{W}{  \gamma}}\times (\apxfac+ \beta) \leq W \\
\Rightarrow &\ceil{\frac{W}{  \gamma}} \leq \frac{W}{(\apxfac+ \beta)}\\
\Rightarrow &\frac{W}{  \gamma} \leq \frac{W}{(\apxfac+ \beta)} +1 \\
\Rightarrow &\gamma \geq (\apxfac+ \beta)\cdot \frac{W}{W+(\apxfac+ \beta)}.
\end{align*}

This gives us the following theorem. 

\begin{theorem}
There exists a $(\apxfac+ \beta)\cdot \frac{W}{W+(\apxfac+ \beta)}$-approximation  algorithm for the \DRTILE problem where  $(\apxfac+ \beta)$ is the approximation factor for the \RTILE problem. The approximation factor of the \DRTILE problem tends to $\frac{3}{2}$ as the value of $W$ is increased. 
\end{theorem}

%% file: 6.multi.tex
\section{The Multidimentional \RTILE Problem} \label{multi}
In Section $3$, the algorithm presented for the \RTILE problem was restricted to two dimensions. In this section, we generalize that algorithm for the $d$-dimensional \RTILE problem, where $d \geq 2$. In the $d$-dimensional \RTILE problem, we are given a $d$-dimensional array of size $n$ in each dimension, containing $0/1$ as entries, and we have to partition the array into $p$ non-overlapping $d$-dimensional tiles such that the maximum weight of a tile in a tiling is minimized. Similarly to Section $3$, we assume that $\frac{p^d}{w(A)}$ is close to $0$ and 
give a $\frac{2d-1}{d}$-approximation algorithm for the $d$-dimensional \RTILE problem. Notice that the approximation ratio converges to $2$ as we increase the value of $d$.

\textbf{Boundaries and Shadows} The definition of the boundaries and their shadows is a generalization of the definitions in Section $3$. The {\em type} of the boundaries in a $d$-dimensional array can be defined analogously. By $[i]$, we define the set of boundaries of dimension $ n\times n\times \ldots \times 1 \times \ldots \times n$, where $i^{th}$ dimension has size $1$.

Let $B_1, B_2, .. B_k \in [i]$ , we define $T_{i} = \sum_{i=1}^kt(B_i)$. Finally $T$ is defined as $min\{T_1, T_2,...,T_d\}$. The following lemma is analogous to Fact $2$.

\begin{lemma}
Let $T = \{T_1, T_2,...,T_d\}$, then the array can be $\frac{2d-1}{d}$-tiled with $T+1$ tiles. 
\end{lemma}

We can estimate the minimal weight of the array using a linear program. The constraints of the linear program have a similar form	as mentioned in Section $3$. In two dimensional problem, each constraint is greater than $1.5L$. In the d-dimensional \RTILE problem, each constraint is greater than $\frac{2d-1}{d}L$, instead of $1.5L$.

\begin{lemma}
Let $T = \{T_1, T_2,...,T_d\}$, then $w(A)>TL$.
\end{lemma}

The proof of this lemma is analogous to Lemma \ref{weight1}. 

\begin{theorem}
There exists a $\frac{2d-1}{d}$-approximation algorithm for the multi-dimensional \RTILE problem assuming $\frac{p^d}{w(A)}$ is negligible. 
\end{theorem}
